\documentclass[prd,showpacs,showkeys,preprintnumbers,amsmath,amssymb,superscriptaddress,floatfix,nofootinbib]{revtex4} 
\usepackage{graphicx,color,dcolumn,booktabs,bm}
\usepackage{longtable,lscape}
\usepackage{txfonts}
\usepackage{overpic}
\usepackage{epsfig}
\usepackage{amssymb}
\usepackage{rotating}
\usepackage{epstopdf}
\usepackage{indentfirst}
\usepackage{feynmf}   
\usepackage{slashed}  
\usepackage{cases}
\usepackage{color}
\usepackage{graphicx,color,dcolumn,booktabs,bm}
\usepackage{cases}
\usepackage{array}
\newcommand{\eq}{\begin{eqnarray}}
\newcommand{\en}{\end{eqnarray}}

\begin{document}

\title{Selected strong decays of 
$\eta(2225)$ and $\phi(2170)$ as $\Lambda \bar\Lambda$ bound states}
\noindent 
\author{Yubing Dong}
\affiliation{
Institute of High Energy Physics, Beijing 100049, P. R. China}
\affiliation{
Theoretical Physics Center for Science Facilities (TPCSF), CAS,
Beijing 100049, People’s Republic of China}
\affiliation{School of Physical Sciences, University of Chinese
Academy of Sciences, Beijing 101408, China}
\author{Amand  Faessler}
\affiliation{
Institut f\"ur Theoretische Physik,  Universit\"at T\"ubingen,
Kepler Center for Astro and Particle Physics,
Auf der Morgenstelle 14, D--72076 T\"ubingen, Germany}
\author{Thomas Gutsche}
\affiliation{
Institut f\"ur Theoretische Physik,  Universit\"at T\"ubingen,
Kepler Center for Astro and Particle Physics,
Auf der Morgenstelle 14, D--72076 T\"ubingen, Germany}
\author{Qifang L\"{u}}
\affiliation{
Institute of High Energy Physics, Beijing 100049, P. R. China}
\affiliation{
Theoretical Physics Center for Science Facilities (TPCSF), CAS,
Beijing 100049, People’s Republic of China}
\affiliation{School of Physical Sciences, University of Chinese
Academy of Sciences, Beijing 101408, China}
\author{Valery E. Lyubovitskij}
\affiliation{
Institut f\"ur Theoretische Physik,  Universit\"at T\"ubingen,
Kepler Center for Astro and Particle Physics,
Auf der Morgenstelle 14, D--72076 T\"ubingen, Germany}
\affiliation{Departamento de F\'\i sica y Centro Cient\'\i fico 
Tecnol\'ogico de Valpara\'\i so-CCTVal, Universidad T\'ecnica
Federico Santa Mar\'\i a, Casilla 110-V, Valpara\'\i so, Chile}
\affiliation{Department of Physics, Tomsk State University,
634050 Tomsk, Russia}
\affiliation{Laboratory of Particle Physics, Tomsk Polytechnic University,
634050 Tomsk, Russia}

\date{\today}

\begin{abstract}

The strong decays of the two resonances $\eta(2225)$ and $\phi(2170)$ are 
discussed for selected decay channels. The two resonances are considered as 
the $\Lambda \bar{\Lambda}$ bound states in the molecular scenario.
The phenomenological hadronic molecular approach is employed for
the calculation of respective decay modes using effective Lagrangians. 
Our results show that the decay modes $\eta(2225)\to K^*K$ 
and $\phi(2175) \to KK$ dominate over
the partial decay widths of $\eta(2275)\to VV 
(\phi\phi, \omega\omega, K^*K^*)$ 
and $\phi(2175)\to VS (\omega\sigma, K^*K^*_0(800), \phi f_0(980))$ 
due to phase space and couplings. 

\end{abstract}



\date{\today}
\maketitle
\section{Introduction}
{\label{introduction}}

Recently, the BESIII Collaboration performed a partial wave analysis
of the decay process $J/\psi \to \gamma \phi \phi$, and confirmed
the existence of the $\eta(2225)$ state,
which has a mass of $2216^{+4+21}_{-5-11}~\rm{MeV}$ and a width of
$185^{+12 +43}_{-14 -17}~\rm{MeV}$~\cite{Ablikim:2016hlu}.
The quantum numbers of $\eta(2225)$ were assigned to be
$I^G(J^{PC}) = 0^+(0^{-+})$. There are only a few theoretical studies on 
$\eta(2225)$ in the literature.
In Refs.~\cite{Li:2008we,Li:2008et} the strong decays of $\eta(2225)$
as a conventional $s \bar s $ state together with its partners were
investigated in the framework of the quark-pair creation model,
and the $4^1S_0$ $s\bar s$ assignment was favored for the $\eta(2225)$ state.
An alternative interpretation of $\eta(2225)$
as a bound state of $\Lambda \bar \Lambda(^1S_0)$ has been proposed
in the one-boson exchange model in Ref.~\cite{Zhao:2013ffn}.
Conversely, the $\phi(2170)$ state with $I^G(J^{PC}) = 0^-(1^{--})$,
denoted previously as $Y(2175)$, has been considered using different
physical interpretations. The mass and width of the $\phi(2170)$ state
are $2180\pm10$~MeV and $83\pm12$~MeV, respectively~\cite{Olive:2016xmw}. 
We also quote a recent result of the BESIII Collaboration~\cite{Ablikim:2017auj} 
for the mass $2135 \pm 8 \pm 9$ MeV and for the width  
$104 \pm 24 \pm 12$ MeV. 
Taking into account information about production and decays of
the $Y(4260)$ state~\cite{Yuan:2008br}, $\phi(2170)$ might
be its strange partner. Possible interpretations include a
traditional $s\bar s$ state~\cite{Ding:2007pc,Shen:2009zze,Wang:2012wa,%
Afonin:2014nya}, hybrid state~\cite{Ding:2007pc,Ding:2006ya},
tetraquark state~\cite{Wang:2006ri,Chen:2008ej,Drenska:2008gr},
$\Lambda \bar \Lambda(^3S_1)$ bound state~\cite{Zhao:2013ffn,Deng:2013aca},
and $\phi K \bar K$ resonance
state~\cite{MartinezTorres:2008gy,GomezAvila:2007ru}.

In the traditional quark model the total decay width of $\eta(2225)$
can be described well by considering it as the $4^1S_0$
state~\cite{Li:2008we}. However, when assigning $\phi(2170)$ as the
$3^3S_1$ or $2^3D_1$ state, then it will result in a much larger decay
width~\cite{Barnes:2002mu,Ding:2007pc,Wang:2012wa} than observed.
Moreover, the very small mass difference between these two states can 
hardly be explained within the quark potential model, in which the mass
of the $4S$ state should be much higher than that of the $3S$ state, even
if the spin fine splitting is taken into account~\cite{Godfrey:1998pd}.
The interpretation of $\phi(2170)$ as the $4^3S_1$ state also causes
the reversal of the fine structure~\cite{Afonin:2014nya}.
Considering that the masses of $\eta(2225)$ and $\phi(2170)$ are very
close to the $\Lambda \bar \Lambda$ threshold, it also seems natural 
that $\eta(2225)$ and $\phi(2170)$ are considered
as the $\Lambda \bar \Lambda(^1S_0)$ and 
$\Lambda \bar \Lambda(^3S_1)$ bound states, respectively~\cite{Zhao:2013ffn}.
Within the one-boson exchange model the mass of the
$\Lambda \bar \Lambda(^1S_0)$ state is slightly higher
than that of the $\Lambda \bar \Lambda(^3S_1)$ state, which is
in good agreement with experimental data~\cite{Ablikim:2016hlu,Olive:2016xmw}.
Besides the mass spectrum, it is natural to examine the strong decay
behavior within the same framework. Note that "deuteronlike" states 
near the respective baryon-antibaryon threshold were originally discussed 
in the context of the nucleon-antinucleon system.
There the notion of so-called quasinuclear $N \bar N$ bound states, weak 
composites of $N\bar N$, and their properties was intensely pursued to explain
resonance structures observed in $N\bar N$ annihilation reactions. For one of 
our contributions to this topic see, for example, Ref.~\cite{Dover:1990kn}.

In this paper, we present a study of selected strong decay modes
of the $\eta(2225)$ and $\phi(2170)$ states. We employ a hadronic 
molecular scenario~\cite{Faessler:2007gv}-\cite{Dong:2017gaw} 
by taking the two resonances as weakly bound states of $\Lambda \bar \Lambda$  
in a phenomenological Lagrangian approach. It should be mentioned that the 
approach is based on the compositeness 
condition~\cite{Weinberg:1962hj}-\cite{Branz:2009cd} ---  
a powerful method in quantum field theory for the study of composite bound 
states (hadrons, glueballs, hybrids, hadronic atoms and molecules, 
multiquark states), which was extensively used 
in Refs.~\cite{Efimov:1993ei}-\cite{Branz:2009cd} 
and~\cite{Faessler:2007gv}-\cite{Dong:2017gaw}. 
In particular, the compositeness condition gives an equation for the 
coupling constant of the bound state with its constituents where the mass of 
the bound state is the input parameter. 
We suppose that our analyses of the $\eta(2225)$ and $\phi(2170)$ strong
decays are useful for running and future experiments.

This paper is organized as follows.
In Sec.~\ref{formalisms} we briefly show our formalism, the calculations
for the couplings of $\eta(2225)\Lambda\bar{\Lambda}$ and
$\phi(2175)\Lambda\bar{\Lambda}$, and the matrix elements for the transitions
of $\eta(2225)\to VV$ (vector-vector mesons), 
   $\eta(2225)\to VP$ (vector-pseudoscalar mesons), 
   $\phi(2175)\to VS$ (vector-scalar mesons), and 
   $\phi(2175)\to PP$ (pseudoscalar-pseudoscalar mesons) 
in the hadronic molecular scenario. 
In Sec.~\ref{results} we present an application of
our approach to the selected strong decays of $\eta(2225)$ and $\phi(2170)$
states. A short summary is given in Sec.~IV.

\section{Approach}
{\label{formalisms}}

In our numerical calculation, we use the following spin-parity quantum numbers 
for the $\eta(2225)$ and $\phi(2170)$ states $J^{PC} = 0^{-+}$ and 
$J^{PC} = 1^{--}$,
respectively. Since the masses of the $\Lambda$ baryon [$I(J^P)=0(1/2^+)$] 
and the $\Lambda\bar{\Lambda}$ system are $1115.683\pm 0.006$ MeV and
about 2232 MeV, respectively, the discussed $\eta(2215)$ 
and $\phi(2175)$ resonances 
are about  $16$ MeV and $57$ MeV below the threshold of
the $\Lambda\bar{\Lambda}$ system. We consider the states $\eta(2225)$
and $\phi(2170)$ as weakly bound states of $\Lambda$ and $\bar{\Lambda}$
in the hadronic molecular scenario. For this purpose we employ our 
phenomenological Lagrangian approach to describe these resonances.
The interaction Lagrangians, describing the couplings of
the $\eta(2225)$ and $\phi(2170)$ $\Lambda\bar{\Lambda}$  
baryonium states with the constituents, read
\eq
{\cal L}_{\eta}(x) &=& g_{\eta} \, \eta(x) \,
\int d^4y \, \Phi(y^2) \,
\bar \Lambda (x + y/2) i\gamma_5 \Lambda(x - y/2)\,,\\
{\cal L}_{\phi}(x) &=&
g_{\phi} \, \phi^{\mu}(x) \,
\int d^4y \, \Phi(y^2) \, \bar \Lambda (x + y/2)
\gamma_{\mu} \Lambda(x - y/2) \,.
\end{eqnarray}
Here $\Phi(y^2)$ is a phenomenological correlation function describing the
distribution of $\Lambda$ and $\bar \Lambda$ constituents
in the $\eta(2225)$ and $\phi(2170)$ states.
To produce ultraviolet-finite Feynman diagrams,
the Fourier transform of the correlation function $\Phi(y^2)$
should vanish sufficiently fast in the ultraviolet region of the Euclidean space.
We use the Gaussian form for the correlation function
\begin{eqnarray}
\tilde\Phi(p^2_E) \doteq \exp(-p^2_E /\Lambda_H^2)\,, \quad H = \eta(2225), \phi(2170)
\end{eqnarray}
where $p_E$ is the Euclidean Jacobi momentum and $\Lambda_H$
is a free size parameter, which has a value of about 1 GeV.

The couplings of $g_{\eta}$ and $g_{\phi}$ with the $\Lambda$ and $\bar{\Lambda}$ 
constituents are calculated from the compositeness
condition (see Refs.~\cite{Weinberg:1962hj}-\cite{Branz:2009cd} 
and~\cite{Faessler:2007gv}-\cite{Dong:2017gaw}) 
\eq
Z_{H} = 1 - \Sigma_{H}^\prime(m^2_H) \equiv 0 \,,
\en
where $\Sigma_H^\prime$ is
the derivative of the mass operator in the case of $\eta(2225)$
and of the transverse part of the mass operator
$\Sigma^T_{\phi(2170)}$ in the case of the $\phi(2170)$ state, respectively. 
Note that the compositeness condition gives the relation between 
the coupling constant $g_H$ of the bound state with their constituents and 
its mass $m_H$. 

The quantity $Z_H^{1/2}$ is the matrix element between 
a physical particle state
and the corresponding bare state. The compositeness condition $Z_H=0$ enables
one to represent a bound state by introducing a hadronic field interacting
with its constituents so that the renormalization factor is equal to zero.
This does not mean that we can solve the QCD bound state equations but we are
able to show that the condition $Z_H=0$ provides an effective and
self--consistent way to describe the coupling of a hadron to its
constituents. 
In particular, the compositeness condition gives an equation for the
coupling constant of the bound state with its constituents where the mass of
the bound state is the input parameter.
One  starts with an effective interaction Lagrangian written
down in terms of quark and hadron variables. Then, by using Feynman rules,
the $S$--matrix elements describing hadron-hadron interactions are given in
terms of a set of quark level Feynman diagrams.

Decomposition of the $\phi(2170)$ mass operator
in the transverse $\Sigma^T_{\phi(2170)}$ and longitudinal
$\Sigma^L_{\phi(2170)}$ parts reads
\eq
\Sigma_{\phi}^{\mu\nu}(p) =
  g_{\perp}^{\mu\nu} \Sigma_{\phi}^T(p^2) 
+ \frac{p^\mu p^\nu}{p^2} \Sigma_{\phi}^L(p^2),
\en
where $g^{\mu\nu}_{\perp}=g^{\mu\nu}-p^\mu p^\nu /p^2$. 
The corresponding Feynman diagrams describing the mass operators of
the $\eta(2225)$ and $\phi(2170)$ states are shown in Fig.~1.

The expressions for the mass operators of $\eta(2225)$
and $\phi(2170)$ are given by
\eq
\Sigma_{\eta}(p^2) &=& g_{\eta}^2 \,
\int \frac{d^4k}{(2\pi)^4 i} \tilde\Phi^2(-k^2) \,
{\rm Tr}\Big[\gamma^5 S_\Lambda(k+p/2) \gamma^5 S_\Lambda(k-p/2)\Big] \,, \\
\Sigma_{\phi}^{\mu\nu}(p) &=&
g_{\phi}^2 \,
\int \frac{d^4k}{(2\pi)^4 i} \tilde\Phi^2(-k^2) \,
{\rm Tr}\Big[\gamma^\mu S_\Lambda(k+p/2) \gamma^\nu S_\Lambda(k-p/2)\Big] \,,
\en
where $S_\Lambda(k) = 1/(m_\Lambda - \not\! k)$ is
the free $\Lambda$ spin-1/2 baryon propagator with $m_\Lambda$ being the mass 
of the $\Lambda$ hyperon. 

The expressions for the coupling constants $g_H$ are given by
\eq\label{coupling_c}
\frac{g_{H}^2}{4\pi^2} &=& \frac{1}{I_H}\,,
\en
where $I_H$ is the structure integral
\eq
I_H &=& \frac{1}{2} \,
\int\limits_0^\infty \, \frac{d\alpha d\beta}{\Delta^3} \, u_H \, e^{-w_H} \,, 
\quad \Delta = 1  + \alpha + \beta  
\en
and 
\eq
w_H &=&
\frac{2m_\Lambda^2}{\Lambda_H^2} \, (\alpha + \beta) -
\frac{m_H^2}{2 \Lambda_H^2}
\, \frac{\alpha+\beta+4\alpha\beta}{1+\alpha+\beta} \,, \nonumber\\
u_{\eta} &=& \frac{m_\Lambda^2}{\Lambda_H^2} (\alpha + \beta + 2 \alpha \beta)
+ \frac{1 + 4 (\alpha + \beta) + 12 \alpha \beta}{2 \Delta}
+ \frac{m_\eta^2}{4 \Lambda_H^2} \,
\frac{(1+2\alpha) (1+2\beta) (\alpha+\beta+4\alpha\beta)}{\Delta^2}
\,, \nonumber\\
u_{\phi} &=& \frac{m_\Lambda^2}{\Lambda_H^2} (\alpha + \beta + 2 \alpha \beta)
+ \frac{1 + 3 (\alpha + \beta) + 8 \alpha \beta}{2 \Delta}
+ \frac{m_\phi^2}{4 \Lambda_H^2} \,
\frac{(1+2\alpha) (1+2\beta) (\alpha+\beta+4\alpha\beta)}{\Delta^2}
\,.
\en
The use of the central values of the $\eta(2225)$ and $\phi(2170)$ masses 
$m_{\eta(2225)} = 2221$ MeV and $m_{\phi()} = 2188$ MeV 
in Eqs.~(\ref{coupling_c}) gives predictions for the 
$g_{\eta(2225)}$ and $g_{\phi(2170)}$ couplings. 

\begin{figure}[htbp]
\includegraphics[scale=0.6]{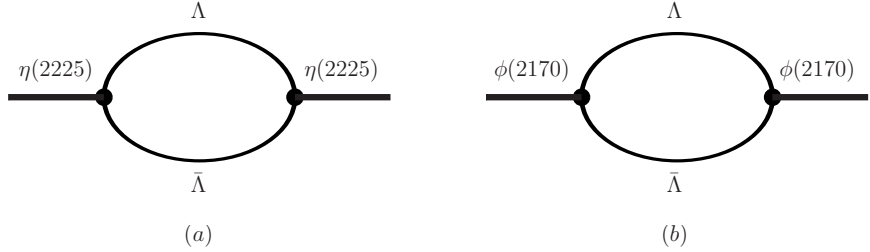}
\vspace*{-0.5cm} \caption{Mass operators of $\eta(2225)$ and $\phi(2170)$.}
\label{mass}
\end{figure}

In this paper we calculate some selected  strong two-body decays
$\eta(2225)\to VV$ and $\phi(2170)\to VS(PP)$, which are described by the 
Feynman diagrams shown in Fig.~2. For the additional hadronic interaction 
vertices the empirical meson-baryon form factors ${\cal F}(x-y)$ are employed. 
Those effective Lagrangians are 
\eq\label{L_KstarLN}
{\cal L}_{K^* \Lambda N}(x) &=& -g_{K^* \Lambda N} 
\, K^{*\mu}(x) \, \bar\Lambda(x) \, \gamma_{\mu} \, \int d^4y \, 
{\cal F}_N(x-y) \, N(y) 
\, + \, {\rm H.c.}\,, \\
{\cal L}_{\omega \Lambda \Lambda}(x) &=&
-g_{\omega \Lambda \Lambda} \, \omega^{\mu}(x) \, 
\bar \Lambda(x) \,  \gamma_{\mu} \, \int d^4y \, {\cal F}_\Lambda(x-y) 
\, \Lambda(y) 
\, + \, {\rm H.c.}\,, \\
{\cal L}_{\phi \Lambda \Lambda}(x) &=& -g_{\phi \Lambda \Lambda} 
\, \phi^{\mu}(x) \,  
\bar\Lambda(x) \, \gamma_{\mu} \, \int d^4y \, {\cal F}_\Lambda(x-y) 
\, \Lambda(y) \, + \, {\rm H.c.}\,, \\
{\cal L}_{a_0(980) \Lambda \Sigma}(x) &=& g_{a_0(980) \Lambda \Sigma} 
\, a_0(x) \, \bar \Lambda(x) \, \int d^4y \, {\cal F}_\Sigma(x-y) 
\, \Sigma(y) + {\rm H.c.}\,, \\
{\cal L}_{K_0^*(800) \Lambda N}(x) &=& g_{K_0^*(800) \Lambda N} \, K_0^*(x) 
\, \bar \Lambda(x) \, \int d^4y \, {\cal F}_N(x-y) \, N(y)  + {\rm H.c.}\,, \\
{\cal L}_{\sigma \Lambda \Lambda}(x) &=& g_{\sigma \Lambda \Lambda} 
\, \sigma(x) \, \bar \Lambda(x) \, \int d^4y \, {\cal F}_\Lambda(x-y) 
\, \Lambda(y) \, + \, {\rm H.c.}\,, \\
{\cal L}_{f_0(980) \Lambda \Lambda}(x) &=& g_{f_0(980) \Lambda \Lambda} 
\, f_0(x) \, \bar \Lambda(x) \, \int d^4y \, {\cal F}_\Lambda(x-y) 
\, \Lambda(y) \, + \, {\rm H.c.} \,, \\
{\cal L}_{K\Lambda N}(x) &=& g_{K \Lambda N} 
\, K(x) \, \bar\Lambda(x) \, i\gamma_5 \, \int d^4y \, 
{\cal F}_N(x-y) \, N(y) 
\, + \, {\rm H.c.}\,. 
\label{L_f0LL}
\en
In the case of vector meson-baryon couplings we restrict to the
minimal coupling --- leading-order contribution in the inverse
baryon mass expansion; i.e., we neglect the nonminimal couplings 
(or ignore the tensor coupling in the $VBB$ interaction as 
in ~\cite{Zhao:2013ffn}).
We fix meson-nucleon couplings using $SU(3)$ symmetry predictions
and phenomenological constraints~\cite{Zhao:2013ffn,Doring:2010ap},
\eq
g_{K^*\Lambda N} &=& - \frac{1}{\sqrt{3}} \,
(2 \alpha_V + 1) g_{\rho NN} \,, \nonumber\\
g_{\omega\Lambda \Lambda} &=&
\frac{2}{3} \, (5 \alpha_V - 2) g_{\rho NN}
\,,\nonumber\\
g_{K_0^*\Lambda N} &=& - \frac{1}{\sqrt{3}} \,
(2 \alpha_S + 1) g_{a_0(980) NN} \,, \nonumber\\
g_{\sigma\Lambda\Lambda} &=& \frac{2}{3} \, (5 \alpha_S - 2) g_{a_0(980) NN}\,, 
\nonumber\\
g_{K\Lambda N} &=& - \frac{1}{\sqrt{3}} \,
(2 \alpha_P + 1) g_{\pi NN} \,, 
\en 
where $\alpha_V = \alpha_S = 1$ and $\alpha_P = 0.4$.
The set of numerical values of the meson-baryon couplings is listed
in Table 1~\cite{Zhao:2013ffn,Doring:2010ap}.
Here we employ the monopole-type form factor $\tilde {\cal F}_B(q^2)$ 
(in momentum space) of the form 
\eq 
{\cal F}_B(x)=\int d^4x e^{-iqx} \tilde{\cal F}_B(q^2)\,, \quad 
\tilde{\cal F}_B(q^2) &=& \frac{\Lambda_B^2-M_B^2}{\Lambda_B^2-q^2} 
\en 
proposed in Ref.~\cite{Cheng:2004ru} and extensively used in 
literature~\cite{Cheng:2004ru}-\cite{Yu:2017zst}  
with $M_B$ being the exchange baryon mass and $\Lambda_B$ being the cutoff
parameter for the exchange momentum. 
According to the discussion
in the literature~\cite{Cheng:2004ru}-\cite{Yu:2017zst}, we choose
$\Lambda_B=M_B+\alpha\Lambda_{QCD}$ with $\Lambda_{QCD} = 220~\rm{MeV}$.
These form factors are necessary to be consistent with the phenomenological
Lagrangians utilized before in~\cite{Zhao:2013ffn}. 

\begin{figure*}[!htp]
\includegraphics[scale=0.5]{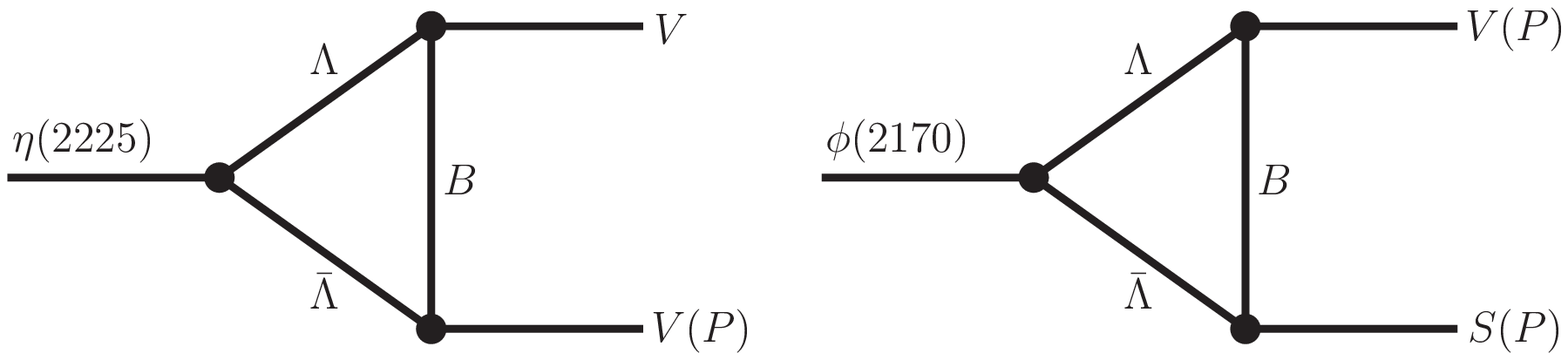}
\vspace*{-1cm} \caption{Feynman diagrams describing
decays $\eta(2225)\to VV(VP)$ and $\phi(2170)\to VS(PP)$.}
\label{feyn1}
\end{figure*}
Now it is straightforward to write down the matrix elements for the discussed 
two-body transition, 
\eq\label{M_inv1}
{\cal M}_{\eta(2225) \to VV}^{\alpha\beta} 
\, \epsilon^*_\alpha(q_1) \, \epsilon^*_\beta(q_2) 
&=& 2 \, \epsilon^*_\alpha(q_1) \, \epsilon^*_\beta(q_2) \, 
g_{\eta} \, g_{V \Lambda B}^2 \, 
\int \frac{d^4k}{(2\pi)^4i} \ \tilde\Phi(-k^2) 
\ \tilde{\cal F}_B^2\Big((k+p/2-q_1)^2\Big) \nonumber\\
&\times& {\rm Tr}\biggl[\gamma^{\alpha} \, S_\Lambda(k+p/2) \, i\gamma^5 \,
S_\Lambda(k-p/2) \, \gamma^\beta
S_B(k+p/2-q_1) \biggr] \nonumber\\
&=& 
\frac{g_{\eta VV}}{m_{\eta(2225)}} \, \epsilon^{\mu\nu\alpha\beta} 
\, q_{1\mu} \, q_{2\nu} \, 
\epsilon^*_\alpha(q_1) \, \epsilon^*_\beta(q_2)\,, \nonumber\\
{\cal M}_{\eta(2225) \to VP}^{\alpha} 
\, \epsilon^*_\alpha(q_1) 
&=& \epsilon^*_\alpha(q_1) \, g_{\eta} \, g_{V \Lambda B} \, g_{P \Lambda B} \, 
\int \frac{d^4k}{(2\pi)^4i} \ \tilde\Phi(-k^2) 
\ \tilde{\cal F}_B^2\Big((k+p/2-q_1)^2\Big) \nonumber\\
&\times& {\rm Tr}\biggl[\gamma^{\alpha} \, S_\Lambda(k+p/2) \, i\gamma^5 \,
S_\Lambda(k-p/2) \, i\gamma^5 
S_B(k+p/2-q_1) \biggr] \nonumber\\
&=& g_{\eta VP} \, q_2^\alpha \, 
\epsilon^*_\alpha(q_1) 
\label{M_inv11}
\en 
for the $\eta(2225) \to VV$ 
and $\eta(2225) \to VP$ 
decays and 
\eq\label{M_inv2}
{\cal M}_{\phi(2170) \to VS}^{\mu\alpha} 
\, \epsilon_\mu(p) \, \epsilon^*_\alpha(q_1) 
&=& \epsilon_\mu(p) \, \epsilon^*_\alpha(q_1) \, 
g_{\phi} \, g_{S \Lambda B} \, g_{V \Lambda B} 
\, \int \frac{d^4k}{(2\pi)^4i} \ \tilde\Phi(-k^2) 
\ \tilde{\cal F}^2_B\Big((k+p/2-q_1)^2\Big) \nonumber\\
&\times&{\rm Tr}\biggl[ S_\Lambda(k+p/2) \, \gamma^\mu 
\, S_\Lambda(k-p/2) \, \gamma^{\alpha}
S_B(k+p/2-q_1) \biggr] \nonumber\\
&=& m_{\phi(2170)} \, 
\biggl( g^{\mu\alpha} g_{\phi VS} \,+\, \frac{q_1^\mu  
\, q_2^\alpha}{m_{\phi(2170)}^2} f_{\phi VS} \biggr)
\, \epsilon_\mu(p) \, \epsilon^*_\alpha(q_1)\,,   \\
{\cal M}_{\phi(2170) \to PP}^{\mu} 
\, \epsilon_\mu(p) 
&=& 2 \, \epsilon_\mu(p) \, 
g_{\phi} \, g_{P \Lambda B} \, g_{P \Lambda B} 
\, \int \frac{d^4k}{(2\pi)^4i} \ \tilde\Phi(-k^2) 
\ \tilde{\cal F}^2_B\Big((k+p/2-q_1)^2\Big) \nonumber\\
&\times&{\rm Tr}\biggl[ S_\Lambda(k+p/2) \, \gamma^\mu 
\, S_\Lambda(k-p/2) \, i\gamma^5 
S_B(k+p/2-q_1) i\gamma^5 \biggr] \nonumber\\
&=& g_{\phi PP} \, (q_1 - q_2)^\mu \, \epsilon_\mu(p) 
\label{M_inv3}
\en
for the $\phi(2170) \to VS$ 
and $\phi(2170) \to PP$ 
decays, 
where $p$ and $q_1$, $q_2$ are the momenta of initial and final particles; 
$g_{\eta VV}$ and $g_{\phi VS}$, $f_{\phi VS}$, $g_{\phi PP}$ 
are dimensionless couplings of $\eta(2225)$ 
and $\phi(2170)$ with final mesons, respectively; 
$\epsilon^*_\mu(p)$, $\epsilon^*_\alpha(q_1)$,  
and $\epsilon^*_\beta(q_2)$ are the polarization vectors of the $\phi(2170)$ 
state and produced vector 
mesons, respectively; $S_B(k)$ is the free spin-1/2 baryon propagator.   
 
Two-body strong decay widths are calculated according to the formulas 
\eq 
\Gamma(\eta(2225) \to VV) &=& \frac{g_{\eta VV}^2}{64 \pi} 
\, m_{\eta(2225)} \, 
\biggl( 1 - \frac{4m_V^2}{m_{\eta(2225)}^2} \biggr)^{3/2}\,, \nonumber\\
\Gamma(\eta(2225) \to VP) &=& \frac{g_{\eta VP}^2}{8 \pi} 
\,\frac{{\bf|q_1|_\eta}^3}{m_V^2}\,, \nonumber\\
\Gamma(\phi(2170) \to VS) &=& 
\frac{g_{\phi VS}^2}{24 \pi} \, {\bf|q_1|_\phi} \, 
\biggl[ 3 + \frac{{\bf|q_1|_\phi}^2}{m^2_V} 
+ \frac{m_{\phi(2170)}^2+m_V^2-m_S^2}{m_{\phi(2170)}^2}
\,\frac{{\bf|q_1|_\phi}^2}{m_V^2}\, R 
+ \frac{{\bf|q_1|_\phi}^4}{m_{\phi(2170)}^2 m_V^2}\,R^2 \biggr]\,, 
\quad R = \frac{f_{\phi VS}}{g_{\phi VS}} \,, \nonumber\\
\Gamma(\phi(2170) \to PP) &=& \frac{g_{\phi PP}^2}{96 \pi} 
\, m_{\phi(2170)} \, 
\biggl( 1 - \frac{4m_P^2}{m_{\phi(2170)}^2} \biggr)^{3/2}\,. 
\en
Here 
${\bf|q_1|_\eta} = \lambda(m_{\eta(2225)}^2,m_V^2,m_P^2)/(2m_{\eta(2225)})$ 
and 
${\bf|q_1|_\phi} = \lambda(m_{\phi(2170)}^2,m_V^2,m_S^2)/(2m_{\phi(2170)})$ 
are the 3-momenta of the decay products in the center of mass frame  
and $\lambda(x,y,z) = x^2 + y^2 + z^2 - 2xy - 2xz - 2yz$ is the K\"allen 
kinematical triangle function.  

\section{Results and Discussions}
{\label{results}}

In Fig.~3 we show the dependence of the couplings $g_H$, $H=\eta(2225),
\phi(2170)$ on the cutoff parameter $\Lambda_H$ [see Eq.~(\ref{coupling_c})].  
When $\Lambda_H$ is varied in the region of (0.8-1.2 GeV), 
the two resulting couplings are not too sensitive to the model parameter 
$\Lambda_H$. 
The variations of the dimensionless couplings are $(3.4\to 3.2)$ and
$(5.8\to 5.1)$, respectively. According to our previous calculations 
in the context of $XYZ$ resonances and to the deuteron system, 
a typical value of $\Lambda_H \sim 1~\rm{GeV}$ is often employed. 
Thus, in this calculation we get $g_{H} = 3.282$ and $5.356$ 
for $\eta(2225)$ and $\phi(2170)$, respectively. 
To make detailed calculations for the decay processes of Fig.~2, the 
couplings of the effective Lagrangians 
in Eqs.~(\ref{L_KstarLN})-(\ref{L_f0LL}) are needed. 
We take these from Refs.~\cite{Zhao:2013ffn,Doring:2010ap} 
as listed in Table I.
 
\begin{figure*}[!htp]
\includegraphics[scale=1.0]{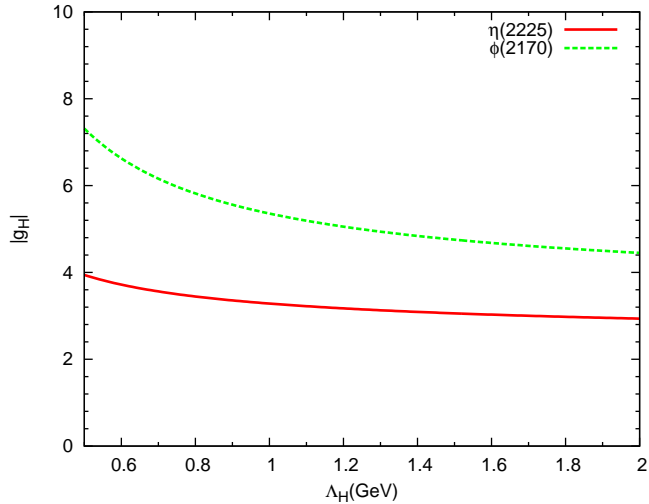}
\vspace*{-0.4cm} \caption{The couplings of $\eta(2225)$ (red-solid line)
and $\phi(2170)$ (green-dashed line) versus the parameter $\Lambda_H$ of 
the correlation function.}
\label{coup}
\end{figure*}

\begin{center}
TABLE I. Effective meson-baryon couplings.

\vspace*{0.1cm}

\def\arraystretch{0.7}
\begin{tabular}{|c|c|c|c|c|c|c|}\hline
&&&&&&\\
  $g_{_{K^*\Lambda N}}$   
 &$g_{_{K\Lambda N}}$   
 &$g_{_{\omega\Lambda\Lambda}}$  
 &$g_{_{\phi\Lambda\Lambda}}$
    &$g_{_{K^*_0(800)\Lambda N}}$          
    &$g_{_{\sigma\Lambda\Lambda}}$   &$g_{_{f_0(980)\Lambda\Lambda}}$  \\
    &&&&&&\\ \hline
    &&&&&&\\
 $- 9.153$  & $- 13.926$ & 10.569  & 5.284   & $- 5.710$  & 6.593  & 3.296 \\
 &&&&&&\\ \hline
\end{tabular}
\end{center}

It should be reiterated that additional phenomenological form factors
$\tilde{\cal F}$ 
in the matrix elements of Eqs.~(\ref{M_inv1}) and (\ref{M_inv2}) 
are introduced, which contain a free parameter $\alpha$. This parameter 
is fixed from data on the total widths of the  
the $\eta(2225)$ and $\phi(2170)$~\cite{Olive:2016xmw}:
$\Gamma_{\eta(2225)} = 185^{+40}_{-20}$ MeV and
$\Gamma_{\phi(2170)} = 83 \pm 12$ MeV.
In particular, an increase of the parameter $\alpha$ 
leads to an increase of the partial widths of $\eta(2225)$ and $\phi(2170)$. 
We compare the sum of the partial modes of the $\eta(2225)$ and $\phi(2170)$, 
which include the dominant channels 
$\eta(2225) \to K^*K$, 
$\eta(2225) \to K^*K^*$, and 
$\eta(2225) \to \omega\omega$ in the case of $\eta(2225)$ and 
$\phi(2170) \to KK$ in the case of $\phi(2170)$, with total widths of 
these states. Using data on the widths of the 
$\eta(2225)$ and $\phi(2170)$ states we found 
that in the case of $\eta(2225)$ the parameter $\alpha$ is constrained as 
$0.91 \leq \alpha \leq 1.08$, while in the case of $\phi(2170)$ 
the parameter $\alpha$ is constrained as $0.85 \leq \alpha \leq 1.0$. 
In both cases the 
lower and upper limits for the $\alpha$ correspond to the lower and upper 
limits for the sum of the partial decays modes, respectively. 
Therefore, taking into account the two above constraints for $\alpha$ 
we finally conclude that from data on the total widths of $\eta(2225)$ and
$\phi(2170)$ the parameter $\alpha$ should be varied in the region 
$0.91 \leq \alpha \leq 1.0$.   

Table II summarizes the numerical results for the partial decay widths of the 
two resonances including the variation of parameter $\alpha$ from $0.91$  
to $1.0$. 
We compare our predictions for the sum of partial widths with  
data for the total widths of $\eta(2225)$ and 
$\phi(2170)$. Also we present a comparison of partial widths   
with available calculations in the $^3P_0$ model 
using the $s\bar s$ interpretation ~\cite{Li:2008we,Wang:2012wa}. 
The much larger decay widths of the 
$\eta(2225)\to K^*K$, $\eta(2225)\to \omega\omega$,  
and $\eta(2225)\to K^*K^*$ channels
compared to $\eta(2225)\to \phi\phi$ are due to the phase space and
particularly to the couplings.  
A similar feature occurs for the $\phi(2170)$ state, for which 
the decay $\phi(2170) \to KK$ dominates over the others 
because of the phase space and relatively big coupling constant 
$g_{K\Lambda N}$. 
We see that for $\eta(2225)$, the $\omega \omega$ channel 
dominates for the $\Lambda \bar \Lambda$ bound state, while it is 
a OZI-forbidden mode within the $s\bar s$ interpretation. 
For $\phi(2175)$, the $^3P_0$ model calculations in the literature 
usually neglect its $SV$ modes and give a rather larger total decay width,   
which disfavor the $s\bar s $ interpretation.                                 
These differences can help us to distinguish the $\Lambda \bar \Lambda$       
bound state and $s\bar s$ interpretation. 

As an independent check of consistency of our results we would 
like to compare our result for the decay width 
$\Gamma(\phi(2170) \to \phi(1020) + f_0(980)) = 0.25 - 0.3$ MeV with data 
$\Gamma(\phi(2170) \to \phi(1020) + f_0(980)) = 0.1  -  1$ MeV, which 
can be extracted using experimental result for  
$\Gamma(\phi(2170) \to \phi(1020) + f_0(980)) \,                           
\Gamma(\phi(2170) \to e^+e^-)/\Gamma_{tot} = (2.3 \pm 0.3 \pm 0.3)$~eV      
and typical values for the branching of ${\rm Br}(\phi(2170) \to e^+ e^-) 
= 10^{-6} - 10^{-5}$ deduced using known data for other $\omega$ states. 

For convenience, in Figs.~4 and 5 we also display the dependence of 
the partial widths of $\eta(2225)$ and $\phi(2170)$ and their sums 
on the parameter $\alpha$ varied in the wide region $0.8 \le \alpha \le 1.5$ 
and compare the total width with the data. Again, one can see that data on the total 
decays of the $\eta(2225)$ and $\phi(2170)$ states give strong 
constraint on the parameter $\alpha$: $0.91 \le \alpha \le 1$.  

\begin{center}
Table 2. Numerical results for the $\eta(2225)$ and $\phi(2170)$ 
decay widths (in MeV). 

\vspace*{0.1cm}

\begin{tabular}{c||c|c|c}\hline
Modes  &\multicolumn{3}{c}{$\Gamma$~(MeV)}\\ \hline
$\eta(2225)$ decay  &This~work 
& $^3P_0$ model within $s\bar s$~\cite{Li:2008we} 
& Data~\cite{Olive:2016xmw} \\ 
\hline 
$K^*K$            &  71.1 $-$  87.3     &  9.1 & $\cdots$ \\
$\phi\phi$        &   1.1 $-$   1.3     & 12.6 &Seen\\
$\omega\omega$    &  53.6 $-$  63.3     &    0 & $\cdots$ \\
$K^*K^*$          &  37.1 $-$  43.7     &  0.5 & $\cdots$ \\ 
Total             & 162.9 $-$ 195.6     & 22.2 & $185^{+40}_{-20}$ \\
\hline\hline
$\phi(2170)$ decay  &This work  &$^3P_0$ model within 
$s\bar s$~\cite{Wang:2012wa} & Data~\cite{Olive:2016xmw} \\ 
\hline
$KK$              &  73.8  $-$  87.7  & $\cdots$ & $\cdots$ \\ 
$\phi f_0(980)$   &   0.25 $-$   0.3  & $<10$ &Seen \\
$\omega\sigma$    &   4.2  $-$   4.9  & $\cdots$ & $\cdots$ \\
$K^*K^*_0(800)$   &   1.8  $-$   2.1  & $\cdots$ & $\cdots$ \\ 
Total             &  80.1  $-$    95  &    &  83 $\pm$ 12 \\
\end{tabular}
\end{center}

\begin{figure*}[!htp]
\includegraphics[scale=1.0]{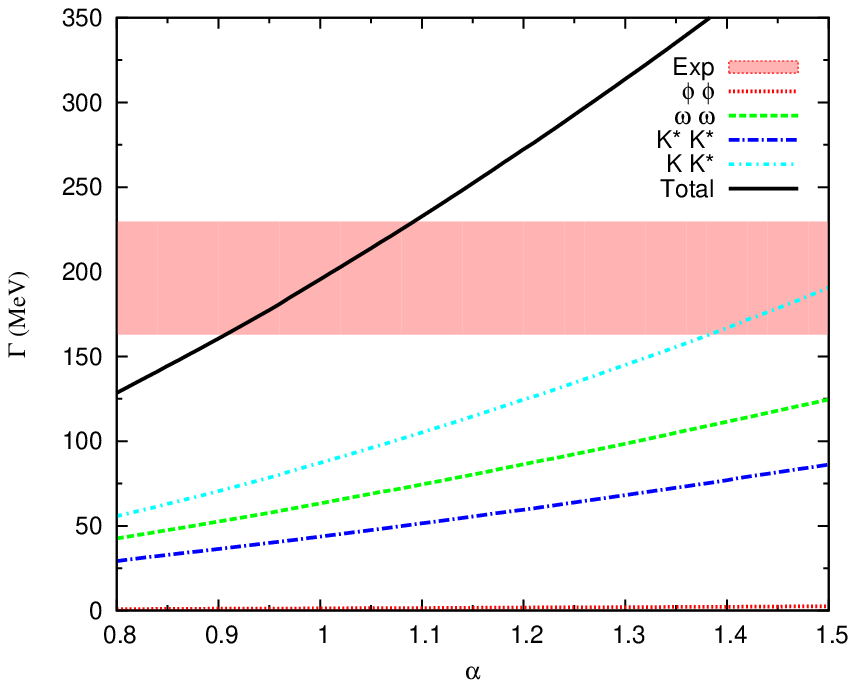}
\vspace*{-0.3cm} \caption{
$\eta(2225)\to VP(VV)$ decays and their sum in dependence on $\alpha$ 
and comparison with data for $\Gamma_{\eta(2225)}$.}
\label{eta}

\includegraphics[scale=1.0]{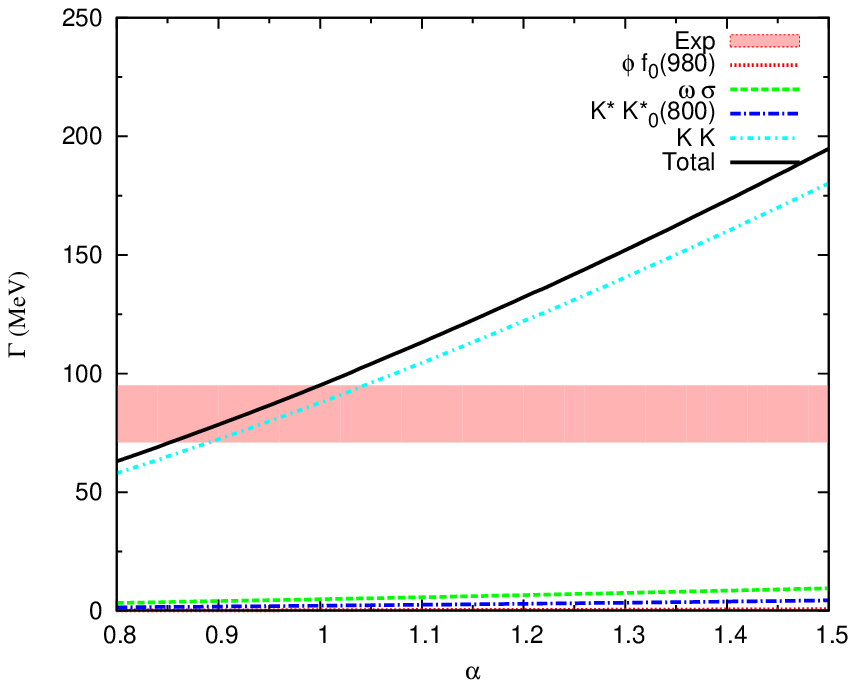}
\vspace*{-0.3cm} \caption{
$\phi(2170)\to VS(PP)$ decays in dependence on $\alpha$ 
and comparison with data for $\Gamma_{\phi(2170)}$.}
\label{phi}
\end{figure*}

\section{Summary}
\label{summary} 

We have employed the hadronic molecular scenario for the two resonances
$\eta(2225)$ and $\phi(2170)$ considering them as weakly bound 
$\Lambda\bar{\Lambda}$ states. A phenomenological effective Lagrangian
approach is applied for some selected partial decay widths.
Our numerical results show that the $\Lambda\bar{\Lambda}$ scenario gives 
a reasonable description of the partial decay widths of the 
$\eta(2225)$ and $\phi(2170)$ states showing that the modes 
$\eta(2225)\to VP$ and $\phi(2170)\to PP$ modes are, respectively, dominant. 
Moreover, together with the study of the mass
spectrum of the two resonances in Ref.~\cite{Zhao:2013ffn},
we conclude that the $\Lambda\bar{\Lambda}$ baryonium interpretations for
the two resonances might be possible. Using data on the total 
widths of the $\eta(2225)$ and $\phi(2170)$ states we derive the constraint 
on the parameter $\alpha$ in the phenomenological form factor controlling  
the off-shell behavior of the exchanged baryon between the produced two final 
mesons: $0.91 \le \alpha \le 1$. For these values of the $\alpha$ parameter 
our predictions for the partial decay widths of $\eta(2225)$ and 
$\phi(2170)$ are shown in Table II. Here we studied selected decay modes 
of the $\eta(2225)$ and $\phi(2170)$ states and included the dominant 
decay modes $\eta(2225)\to VV$ and $\phi(2170)\to PP$. 
There are, of course, many other channels, such
as $\eta(2225)\to N\bar{N}, PV, PS$ and 
$\phi(2170)\to N\bar{N}, PV, SS$, 
which contribute a full coupled-channel calculation 
and will be studied elsewhere.

\begin{acknowledgments}

We thank Bing-Song Zou, Fei Huang, and Yin Huang for
useful discussions. This work is
supported by the National Natural Sciences Foundations of China
under the Grants No. 11475192, No. 11565007, No. 11635009, 
No. 11705056, and No. 11521505 and by the fund of Sino-German CRC 110 
"Symmetries and the Emergence of Structure in QCD"
project by NSFC under Grant No. 11621131001,
China Postdoctoral Science Foundation under Grant No. 2016M601133.
This work was funded by the German Bundesministerium f\"ur Bildung
und Forschung (BMBF) under Project 05P2015 - ALICE at High Rate (BMBF-FSP 202):
``Jet- and fragmentation processes at ALICE and the parton structure
of nuclei and structure of heavy hadrons,'' 
by CONICYT (Chile) PIA/Basal FB0821, by Tomsk State University Competitiveness
Improvement Program and the Russian Federation program ``Nauka''
(Contract No. 0.1764.GZB.2017) and 
by Tomsk Polytechnic University Competitiveness 
Enhancement Program (grant No. VIU-FTI-72/2017).   
Y. B. D. also thanks the support from Alexander von Humboldt
foundation and the hospitality of the Institute of Theoretical Physics,
Tuebingen University, Germany.

\end{acknowledgments}

\end{document}